\documentstyle[12pt,a4wide]{article}

\input{epsf.tex}

\def\CARRO{{\cal C}}
\def\De{\Delta}
\def\Im{{\rm Im}\,}
\def\IK#1{{\cal K}_{#1}}
\def\la{\lambda}
\def\Re{{\rm Re}\,}

\begin{document}

\begin{titlepage}
\renewcommand{\thefootnote}{\fnsymbol{footnote}}
\makebox[2cm]{}\\[-1in]
\begin{flushright}
\begin{tabular}{l}
CERN--TH/95--25\\
UAB--FT--360\\
hep-ph/9502338\\
February 1995
\end{tabular}
\end{flushright}
\vskip0.4cm
\begin{center}
{\Large\bf Next-to-leading Order Radiative Corrections\\[4pt]
to the Decay $b\to ccs$}

\vspace{1.5cm}

E. Bagan$^1$,
Patricia Ball$^2$, B. Fiol$^1$ and P. Gosdzinsky$^1$

\vspace{1.5cm}

$^1${\em Grup de F\'\i sica
Te\`orica, Dept.\ de F\'\i sica and Institut de F\'\i sica d'Altes
Energies, IFAE, Universitat Aut\`onoma de Barcelona,
E--08193 Bellaterra (Barcelona), Spain}\\[0.5cm]
$^2${\em CERN, Theory Division, CH--1211 Gen\`{e}ve 23, Switzerland}

\vspace{4cm}

{\bf Abstract:\\[5pt]}
\parbox[t]{\textwidth}{
We calculate the complete ${\cal O}(\alpha_s)$ corrections to the quark
decay $b\to ccs$ taking full account of the quark masses, but neglecting
penguin contributions. For a c to the b quark mass ratio $m_c/m_b=
0.3$ and a strange quark mass of $0.2\,$GeV, we find that the next-to-leading
order (NLO) corrections increase $\Gamma(b\to ccs)$ by $(32\pm 15)\%$
with respect to the leading order expression, where the uncertainty is
mostly due to scale- and scheme-dependences. Combining this
result with the known NLO and non-perturbative corrections to other B
meson decay channels we obtain an updated value for the semileptonic
branching ratio of B mesons, $B_{SL}$, of $(12.0\pm 1.4)\% $ using pole quark
masses and $(11.2\pm 1.7)\% $ using running $\overline{\mbox{MS}}$ masses.
}

\vfill
{\em Submitted to Phys.\ Lett.\ B.}
\end{center}
\end{titlepage}
\setcounter{footnote}{0}
\newpage

{\large\bf 1.} Owing to the newly developed tool of expanding in the
inverse heavy
quark mass \cite{PLB}, the theoretical description of weak inclusive decays
of heavy mesons now rests on more solid ground than ever. Since in
such decays the energy release is large compared to the masses of
the final state particles, the process takes place essentially at small
distances and, in leading order in the heavy quark expansion (HQE), is
described by the underlying quark decay.
Hadronic corrections only enter at second order in the HQE
and should be $\sim 1\,{\rm GeV}^2/m_b^2$, which is around
$5\%$ for B decays with a b quark mass $m_b\approx 5\,$GeV.
Thus the accuracy of
theoretical predictions for hadronic quantities like, say, the
semileptonic branching ratio is not so much limited by our
necessarily incomplete knowledge of (non-perturbative)
hadronic matrix elements, but rather controlled by our knowledge of
{\em perturbative} corrections to the free quark decay.

This issue has recently attracted much attention in
connection with the summation of certain terms of the perturbative series,
namely the asymptotically leading ones of order $\alpha_s^{n+1}\beta_0^n$
\cite{summation}. Although this program can straightforwardly be applied to
semileptonic B decays\footnote{Cf.\ \cite{luke} for an explicit
calculation of the $\alpha_s^2\beta_0$ term in the decays $b\to u e \nu$ and
$b\to c e \nu$.} \cite{prep}, there are  severe problems (both technical and
conceptual ones) with applying it
to nonleptonic channels, so that in this letter we
only deal with the first order radiative corrections.

Until recently, full ${\cal O}(\alpha_s)$ corrections were only
known for the semileptonic decay $b\to c e\nu$ \cite{HP,nir} and for $b\to c
\tau \nu$ \cite{HP}.
Although it is known
that the exchange of gluons between quarks of unequal masses can yield
big effects (cf.\ the extreme case of an infinitely heavy quark
investigated in Ref.\ \cite{BG92}), finite c quark mass
effects in the ${\cal O}(\alpha_s)$ corrections to
the nonleptonic decay $b\to c u d$ and a rough
estimate\footnote{See also \cite{vol1}.} of c quark mass effects
in $b\to ccs$ have only recently been obtained~\cite{tum67}.
In Ref.\ \cite{B42} part of the effects of finite charm
and strange quark masses in the radiative corrections to $b\to ccs$ were taken
into account, based on the calculation done in Ref.\ \cite{HP}.
In this letter, we complete the calculation
of finite quark mass effects in the ${\cal O}(\alpha_s)$
corrections to $b\to ccs$, neglecting
penguin corrections. We exploit this result to give an updated prediction
for the
semileptonic branching ratio $B_{SL}$ of B mesons.

\bigskip

{\large\bf 2.} In calculating the decay rate $\Gamma(b\to ccs)$, we start
from its representation as the imaginary part of the relevant
forward-scattering amplitude:
\begin{equation}\label{eq:FSA}
\Gamma(b\to ccs) = \frac{1}{m_b}\,{\rm Im}\,i\!\int\!\!d^4x\,\langle\, b\,|\,
T\,{\cal L}_W^{\Delta C=2}(x){\cal L}_W^{\Delta C=2}(0)\,|\,b\,\rangle.
\end{equation}
${\cal L}_W^{\Delta C=2}$ is the effective Lagrangian that describes the
decay process in the limit of an infinite W boson
mass and to first order in the weak coupling. It can be written as
\begin{equation}
{\cal L}_W^{\Delta C=2} = {}-\frac{G_F}{\sqrt{2}}\,V_{cs}^*\,V_{cb}\,
\sum_{i=1}^6c_i(\mu){\cal O}_i(\mu).
\end{equation}
In this letter we conform to the notation of \cite{buras}, where the operators
${\cal O}_1$ and ${\cal O}_2$ denote current-current operators with a
colour-non-singlet and colour-singlet structure, respectively, whereas the
remaining operators are due to the admixture of penguin contributions.
The $c_i$ are perturbatively calculable  short-distance coefficients that
describe the physics between the scale of the W boson and the characteristic
hadronic scale of the process, which is of the order of the b quark mass.
If we neglect the effects of strong interactions, then only ${\cal O}_2$ has a
non-vanishing Wilson-coefficient. In next-to-leading order (NLO), both the
coefficients $c_i$ and the operators ${\cal O}_i$ depend on the scheme used
to deal with $\gamma_5$. To that accuracy, the decay rate
$\Gamma(b\to ccs)$ can be written as
\begin{eqnarray}
\Gamma(b\to ccs) & = & \frac{G_F^2m_b^5}{64\pi^3}\,|V_{cb}|^2
|V_{cs}|^2\,{\rm PH}(x_c,x_c,x_s)
\;\sum\limits_{i=1}^6\sum_{j=1}^i\,f_{ij}c_i(\mu)c_j(\mu)d_{ij}
\label{eq:deffij}\\
& \equiv & \frac{G_F^2m_b^5}{64\pi^3}\,|V_{cb}|^2 |V_{cs}|^2\,
{\rm PH}(x_c,x_c,x_s)\kappa(x_c,x_s,\mu)\,K(x_c,x_s,\mu),
\end{eqnarray}
where
$d_{ij}\equiv 1+
r_{ij}\,[\alpha_s(m_W)-\alpha_s(\mu)]/{\pi} + k_{ij}\,
\alpha_s(\mu)/{\pi}$.
The function $\kappa$ is defined in such a way as to contain
the LO effects, whereas
the product $\kappa K$ covers the complete NLO terms.
In Eq.\ (\ref{eq:deffij}) the $c_i$ denote the {\em leading order}
Wilson-coefficients, so that both the $r_{ij}$ and the $k_{ij}$ are
scheme-independent. PH is the tree-level phase-space factor given by
\begin{equation}
{\rm PH}(x_1,x_2,x_3) = 12\hskip-15pt\int\limits_{(x_2+x_3)^2}^{(1-x_1)^2}
\hskip-5pt
\frac{ds}{s}\,(s-x_2^2-x_3^2)(1+x_1^2-s)w(s,x_2^2,x_3^2)w(s,x_1^2,1)
\end{equation}
with
\begin{equation}\label{eq:defw}
w(a,b,c)=(a^2+b^2+c^2-2ab-2ac-2bc)^{1/2}.
\end{equation}
The arguments of the phase-space factor are ratios of the quark masses,
$x_c = m_c/m_b$ and $x_s=m_s/m_b$. The weight functions
$f_{ij}$ are tabulated in Table 1; they depend on\footnote{Note that we have
corrected a sign error in Ref.\ \cite{B42} in all the terms
containing $f$. We are grateful to G. Buchalla for pointing out this mistake.}
\begin{equation}
f = \frac{1}{{\rm PH}(x_c,x_c,x_s)}\int\limits_{(x_c+x_s)^2}^{(1-x_c)^2}
\hskip-10pt
ds\,\frac{6x_c^2}{s^2}\,w(s,x_c^2,x_s^2)w(1,s,x_c^2)(s+x_s^2-x_c^2)
(1+s-x_c^2).
\end{equation}
$f$ describes the interference of operators with Dirac structure
$(V-A)\otimes (V+A)$ with those of the form $(V-A)\otimes (V-A)$
and vanishes for zero final state
quark masses. The coefficients $r_{ij}$ can be obtained from Ref.\
\cite{buras}. In
particular, we find
\begin{eqnarray}
\lefteqn{\sum\limits_{i=1}^2
\sum\limits_{j=1}^if_{ij}c_i(\mu)c_j(\mu)r_{ij}=}\nonumber\\
& & \frac{10863-1278n_f+80n_f^2}{162\beta_0^2}\,\left[\frac{\alpha_s(m_W)}
{\alpha_s(\mu)}\right]^{4/\beta_0}-\frac{15021-1530n_f+80n_f^2}{162\beta_0^2}
\left[\frac{\alpha_s(m_W)}{\alpha_s(\mu)}\right]^{-8/\beta_0}\hskip-25pt,
\end{eqnarray}
where $\beta_0=11-2n_f/3$ is the lowest order coefficient of the QCD
$\beta$-function; in our case we have $n_f=5$ quark flavours. $k_{11}$
and $k_{22}$
were already given in Ref.\ \cite{B42}; $k_{12}$ can be obtained from the
diagrams shown in Fig.~1 as\footnote{We use the same notations as in Ref.\
\cite{tum67}.}
\begin{equation}
k_{12} = k_{22} + \frac{2}{3}\,(H_e+B) \mbox{\rm\ with\ }
H_e\,{\rm PH}(x_c,x_c,x_s) = \frac{768\pi^5}{g_s^2m_b^6}\,{\rm Im}
({\rm VI} + {\rm VIII} + {\rm X} + {\rm X}^\dagger + {\rm XI} +
{\rm XI}^\dagger).
\end{equation}
Here $B$ is a scheme-dependent constant that removes the scheme-dependence of
$H_e$; in na\"{\i}ve dimensional regularization (see below) one finds $B=11$
\cite{buras}. Note that $H_e$
is independent of the definition of the quark mass, which only affects $k_{11}$
and $k_{22}$ through the self-energy diagrams.

\bigskip

{\large\bf 3.} In the following, we present our results in the limit of
vanishing strange quark mass (as will be discussed below, their dependence
on this parameter is small); the full results are available from the
authors as a Mathematica file. We have checked that all formul\ae\  coincide
with
the corresponding ones in Ref.\ \cite{tum67} when the appropriate limit is
taken.\footnote{Note a misprint in Eq.\ (C.9) in Ref.\ \cite{tum67}: the
factor $(2m_c^2+s)$ should read $(m_c^2+2s)$.}

Without going into too many details, we
present first a short outline of the method of calculation of $H_e$ which is
described at length in Ref.\ \cite{tum67}. In calculating the imaginary
parts of the diagrams of Fig.~1, we use $\overline{{\rm MS}}$ subtraction and
control the ultraviolet divergences through di\-men\-sio\-nal
regularization with an anticommuting $\gamma_5$, often referred to as
na\"{\i}ve dimensional regularization (NDR). NDR is applicable if one uses
Fierz-transformations to relate diagrams with closed fermion loops,
which are ambiguous in NDR, to diagrams which are well-defined in
NDR. As shown in Ref.\ \cite{BW90}, Fierz-transformations are only valid
diagram by diagram  with a correct choice of the so-called
evanescent operators. We have verified that in the limits $m_c, m_s\to 0$
our procedure yields the same results as obtained in other schemes
\cite{ACMP81}. Technically, we calculate the imaginary parts of the
forward-scattering amplitudes by applying Cutkosky
rules. We regularize intermediate infra-red singularities by introducing small
quark and gluon masses, denoted by $\rho$ and $\lambda$ respectively, which
allows phase-space
integration to be done in four dimensions.

For the sake of compactness in displaying the formul\ae, the
square masses of the heavy quarks, c, b, are denoted by $c$, $b$.
In the same spirit, we define $\De=(\sqrt{b}-\sqrt{c})^2$; $w=w(b,c,t)$ (or
$w=w(p_1^2,p_2^2,l^2)$ in Eqs.~(\ref{CARRO}), (\ref{Kpm}) below), where
$w$ was defined in Eq.\ (\ref{eq:defw}),
and similarly $v=w(c,c,t)/t$. Finally, we omit the arguments of the
functions. Our results are written in terms of the functions
$A$, $B$, $C$, $\tilde B$, $\tilde K$, which were defined in
Ref.~\cite{tum67}, Eqs.~(A.1)--(A.4), with $M^2\equiv p_1^2$ and $\mu^2\equiv
p_2^2$. Hence, throughout this letter we consider the
external square momenta $p_1^2$, $p_2^2$ and $(p_1+p_2)^2$ to be the
natural arguments of those functions. The integrals
can be computed following standard techniques and be expressed in terms
of logarithmic and dilogarithmic functions. The final analytic expressions
for $A$, $B$, $C$, $\tilde B$, $\tilde K$ are
rather involved; we will give them elsewhere along with details of the
calculation.
The functions $\IK j$, $j=0,1,\dots, 7$, denote certain phase-space integrals
defined as
\begin{eqnarray}
(\IK0,\dots,\IK7) & = & \int \mbox{LIPS}(p_1,p_2,k)
\left(1,{1\over 2p_1 k +\la^2}, p_1 k,{p_2 k\over 2p_1 k +\la^2},
{1\over \la^2-2k l},{p_2 k\over \la^2-2k l},\right.\nonumber\\
& & \phantom{\int \mbox{LIPS}(p_1,p_2,k)}
\left. {1\over (\la^2-2kl)(2p_2 k+\la^2)},
{1\over (2p_1k+\la^2)(2p_2k+\la^2)}\right),
\end{eqnarray}
where $l=p_1+p_2+k$. 
$\IK0 $,\dots, $\IK5 $, can be obtained from the results given
in Ref.\ \cite{Pietsch}, whereas the calculation of $\IK6$ and $\IK7$
requires some effort. The prime symbol ($'$), as in $\CARRO'$ or $\IK2'$
below, will always denote the replacement
$p_1\leftrightarrow p_2$. After a tedious calculation, one finds
\begin{eqnarray}
(\IK6,\IK7)=
{\pi ^2\over 4 l^2} \left(-\CARRO, \left[\CARRO
+ \CARRO' \right]\right),
\end{eqnarray}
where $\CARRO$ is given by
\begin{eqnarray}
\CARRO&=&
-  2 \ln{m_1 K_+ +m_2 \over m_1}  \ln  {K_+ l^2 \over w (K_+ -1)}
-2 L_2 \left( {K_+ -1 \over K_+ +m_2 / m_1}  \right) \nonumber \\
&-& 2\ln {m_1 K_- +m_2 \over m_1} \ln  {K_+ - K_- \over 1 -K_-}
+{5\over 2} L_2 \left( {K_- -K_+ \over K_- +m_2/m_1 }  \right)
- 2 L_2 \left( {K_- -1 \over K_- + m_2/m_1 } \right) \nonumber \\
&+&  2 \ln {m_2 \over  m_1} \ln K_+
-2 L_2 \left( { -m_1 K_+ \over m_2} \right)
+2 L_2  \left( {- m_1 \over m_2} \right) + \ln  {l^2 \over m_1^2}
\ln {l^2\over l^2-(m_1 +m_2)^2}   \nonumber \\ &+ &
L_2  \left( {l^2-(m_1+m_2)^2 \over l^2} \right)  +
\ln  {K_- +m_2/m_1 \over K_+ +m_2/m_1 }
\ln  {\lambda \over \sqrt {l^2}} -
{1 \over 2} L_2  \left( {K_+ -K_- \over K_+ +m_2/m_1} \right).
\label{CARRO}
\end{eqnarray}
In the above formula, we have introduced the obvious notation
$\sqrt{p_j^2}=m_j$. The functions $K_\pm$ are given by
\begin{eqnarray}
K_\pm =
{l^2-p_1^2-p_2^2 \pm w\over 2 m_1 m_2}.
\label{Kpm}
\end{eqnarray}
Now that we have calculated the ${\cal K}_i$ phase-space integrals,
we may
give the imaginary parts of the relevant diagrams, where we denote
the sum of all $j$-particle cuts for a given diagram by the
superscript ${}^{(j)}$. To start with, we find for diagram VI
\begin{equation}
\Im {\rm VI}^{(j)} =  {1 \over 8 \pi b}
\int_{4c}^{b} {\rm d}t\;
(b-t)^2 \left[
b
 \rho^{(j-1)}_1  -2
 \rho^{(j-1)}_2 \right],
\label{imVI}
\end{equation}
where the spectral densities $ \rho _1^{(j)}$ and $ \rho _2^{(j)}$ are given
by
\begin{eqnarray}
 \rho _1 ^{(2)}&=& \Re {g^2 v\over 24 \pi ^4 t}
\left\{ \phantom{1\over 2}   \hspace{-3mm}
\nonumber
( 4c-t)[t
(A +B)+2(t+c)\tilde B]
 - 2(t+2c)
\left( C+{1\over 2} \right)
+  (t^2-4c^2 )
 \tilde K
  \phantom{1\over 2} \hspace{-3mm}\right\},
\nonumber\\
 \rho _2 ^{(2)}   &=&  \Re
{g^2 v\over 48 \pi ^4 }
\left\{ \phantom{1\over 2}    \hspace{-3mm}
2( t - c )
 [t(A+B)+2C+1-(t-2c)\tilde K ]
+  (4 t^2 - 15ct +2c^2)
 \tilde B \phantom{1\over 2} \hspace{-3mm}\right\},
\nonumber\\
 \rho _1 ^{(3)}  & = &  {g^2 \over 6 \pi ^6 t^2}
\left\{ \phantom{1\over 2}    \nonumber \hspace{-3mm}
t (t^2 -4c^2) \IK7
- 2t(t + c ) \IK1
-  2 t \IK0 +8 c \IK3
+  8 \IK2
\phantom{1\over 2} \hspace{-3mm} \right\},\nonumber\\
 \rho _2 ^{(3)}  & = & {g^2 \over 12 \pi ^6 t}
\left\{ \phantom{1\over 2}   \hspace{-3mm}
- 2t( t - c )  (t - 2c) \IK7 +2t(2 t - c ) \IK1
  + t \IK0 - 4 c \IK3  -  4 \IK2
\phantom{1\over 2} \hspace{-3mm} \right\}.
\end{eqnarray}
The external momenta in
$A$, $B$, $C$, $\tilde B$, $\tilde K$ satisfy $p_1^2=p_2^2=c$,
$(p_1+p_2)^2=t$,
whereas
in~$\rho_i^{(3)}$, $t\equiv(p_1+p_2+k)^2$.
For the remaining diagrams, we obtain:
\begin{eqnarray}
\Im {\rm VIII}^{(3)}  &=&  \Re {-g^2 \over 32 \pi ^5 b}
\int_{c}^{\De}
{{\rm d}t \over t} (c-t)^2
(t - c - b) w \nonumber \\
 &\times&\left\{ \phantom{1 \over 2} \hspace{-3mm} \right.
 2t A +2 t
B - 2(c-t)
{\tilde B} +  8
\left( C + {1\over 16} \right) + \left. (c-t )
{\tilde K}
\phantom{1 \over 2} \hspace{-3mm} \right\},\label{VIII3}\\
\Im {\rm VIII}^{(4)} & = &{g^2 \over 8 \pi ^7 b}
\int_{c}^{\De} {\rm d}t\,
  (t - c - b) w\,\left\{
\phantom{1 \over 2} \right.\hspace{-4mm}\IK0- (t - c)  \IK1
-t \IK1'+ \left. (t - c)^2
\IK7 \phantom{1 \over 2} \hspace{-3mm} \right\}
, \label{VIII4} \\ 
\Im [{\rm X}+{\rm X}^\dagger]^{(3)} & = & {g^2 \over 32 \pi ^5b}
\int_{c}^{\De}
{{\rm d}t \over t}  (t-c)^2 w
\,\left\{ \phantom{1 \over 2} \right. \hspace{-5mm}
(b+c-t)\left[2t(A+B)+8C+{1\over2}\right.\nonumber \\
&+&\left.\phantom{1 \over 2} \hspace{-5mm} (b+c-t)\tilde K \right]
- 2 (t^2 - 2 t c - 2 t b + c^2 + b^2)
{\tilde B} \left. \phantom{1 \over 2}\hspace{-3mm}
\right\}, \label{X3}\\
%
\Im [{\rm X}+{\rm X}^\dagger]^{(4)} & =&{g^2 \over 8 \pi ^7}
\int_{c}^{\De} {{\rm d}t \over t}
(t-c)^2
\left\{ \phantom{1 \over 2}   \hspace{-3mm}
 (t -c - b)^2
\IK6
+ \IK0 - (t - b)
\IK1  -  (t - c)
\IK4 \phantom{1 \over 2} \hspace{-3mm}\right\},\nonumber\\[-10pt]
& & \label{X4}\\[-10pt]
\Im [{\rm XI}+{\rm XI}^\dagger]^{(3)}&=&
{g^2\over192\pi^5 b}\int_{4c}^b{{\rm d}t\over t}\, (b-t)^2 v
\left\{
(t+2c)b\left[t(A+4B)
+(t+2b)\tilde B\right.\right.\nonumber\\
&-&2C-1
-\left. (b-t)\tilde K\right]\nonumber\\
&-&\left. 2 t(t-c)\left[
(t+b)(A+B)-(b-t)(2\tilde B-\tilde K )+2C+1
\right]
\right\}, \label{XI3}\\
\Im [{\rm XI}+{\rm XI}^\dagger]^{(4)}&=&
{g^2\over48\pi^7}\int_{4c}^b{{\rm d}t\over t}\,  v
\left\{
(t+2c)\left[t\IK0+b(t-b)\IK1+2\IK2'+b^2\IK4-2b\IK5\right.\right.\nonumber\\
&-&\left.\left. b(t-b)^2\IK6
\right]
+2t(t-c)\left[(t-b)\IK1+t\IK4-(t-b)^2\IK6
\right]
\right\}.
\label{XI4}
\end{eqnarray}
In Eq.\ (\ref{VIII3}), the arguments of $A$, $B$, $C$,
$\tilde B$, $\tilde K$ satisfy $p_1^2=c$, $p_2^2=\rho^2\to0$. As
mentioned above, $\rho$ (and also $\lambda$) regularizes the infra-red
singularities arising in
intermediate steps of the calculation. They cancel upon addition
of the 3- and 4-particle cut contributions to each diagram.
In Eq.\ (\ref{X3}), we set
$p_1^2=c$, $p_2^2=b$. Finally, in Eq.\ (\ref{XI3}), we have
$p_1^2=\rho^2\to0$, $p_2^2=b$,
and in all three equations $(p_1+p_2)^2=t$. In Eq.\ (\ref{VIII4}),
the arguments
of
the functions $\IK j$ are $p_1^2=\rho^2\to0$, $p_2^2=c$ and
$l^2=t$. In Eq.\ (\ref{X4}), $p_1^2=c$, $p_2^2=t$ and $l^2=b$.
Finally, in Eq.\ (\ref{XI4}), $p_1^2=\rho^2\to0$,
$p_2^2=t$ and $l^2=b$.

The numerical results of our calculation are presented in
Table 2, namely
$k_{12}$ as a function of the charm quark mass for zero strange quark mass
and $m_s=0.2\,$GeV, respectively. For comparison and completeness we likewise
give the coefficients $k_{11}$ and $k_{22}$ referring
 to the on-shell definition
of quark masses. The table shows also the leading order correction $\kappa$
and the ratio $\Gamma_{NLO}/\Gamma_{LO}=K$. In the latter quantity, we
have estimated
the unknown NLO penguin contributions very conservatively by
assuming $0<d_{ij}<2$, which
corresponds to $|k_{ij}|<15$ for $\alpha_s=0.2$. The other input parameters
are given in the table caption. The effect of a finite value of the strange
quark mass is tiny for the rate (though appreciable for the NLO corrections
after dividing out the phase-space factor), and less than 5\% for $x_c\leq
0.4$, which is less than the estimated uncertainty from the unknown NLO
penguin contributions. Using pole masses and a renormalization scale
$\mu = m_b$, we thus observe that $\Gamma(b\to ccs)$ increases by $(32\pm 7)\%$
through NLO
corrections for a reasonable choice of quark masses
$x_c = 0.3$. If we allow the renormalization scale to vary in
the range\footnote{We conform here to the conservative choice of
``characteristic scales'' that is preferred if one follows standard
renormalization group improvement arguments, where large logarithms of
type $\ln m_b^2/\mu^2$ are to be avoided. The results obtained from
summing the asymptotically leading part of the perturbative series seem,
however, to indicate a lower scale, at least
in semileptonic decays, cf.\ \cite{summation,luke,prep}. It remains
to be seen if those scales necessarily have to coincide or not.}
$m_b/2 < \mu < 2m_b$ and take the uncertainty in $x_c$ to be $\pm 0.05$,
we obtain $\Gamma_{NLO}/\Gamma_{LO} = 1.32\pm 0.15$.

So far we have used the on-shell definition of the quark mass. However,
far from being com\-pul\-so\-ry, this definition most likely introduces
artificially large higher order perturbative corrections (cf.\ \cite{prep}).
It is therefore most instructive to eliminate the pole mass in favour of
an off-shell renormalized mass, such as, e.g. the $\overline{\mbox{MS}}$ mass.
As discussed in Ref.\ \cite{BN}, this amounts to the replacement
\begin{equation}
m_b^5\,{\rm PH}(x_c,x_c,0) \longrightarrow \bar{m}_b^5\,
{\rm PH}(\bar{x}_c,\bar{x}_c,0)\left\{1+
\frac{\alpha_s}{\pi}\left(\frac{20}{3}-5\ln\,\frac{\bar{m}_b^2}{\mu^2}-2
\bar{x}_c \ln \bar{x}_c\, \frac{d\ln{\rm
PH}(\bar{x}_c,\bar{x}_c,0)}{d\bar{x}_c} \right)\right\}
\end{equation}
in the decay rate,
where $\bar{x}$ denotes a running quantity evaluated at the scale
$\mu$. We then obtain\footnote{Note that $x_c = 0.30\pm 0.05$ translates
to $\bar{x}_c(\mu=m_b) = 0.28\pm 0.05$.} $\Gamma_{NLO}/\Gamma_{LO} =
1.2\pm 0.4$,
which indicates that the uncertainty due to unknown higher order corrections
is appreciable. We shall come back to this point in the next section.

\bigskip

{\large\bf 4.} With the results for $\Gamma(b\to ccs)$ in hand, we
are ready to give an updated value for the semileptonic branching ratio
$B_{SL}$ of B mesons defined as
\begin{equation}\label{eq:BRSL}
B_{SL} \equiv  \frac{\Gamma(B\to X e\nu)}{\sum_{\ell =
e,\,\mu,\,\tau}\!\Gamma(B\to X\ell\nu_{\ell}) + \Gamma(B\to
X_c) + \Gamma(B\to X_{c\bar c})+\Gamma({\rm rare\ decays})}.
\end{equation}
Performing an expansion in the inverse b quark mass, it is possible to show
\cite{PLB} that the inclusive decay rate of a B meson into a final state X
coincides with that of the underlying b quark decay up to
corrections of order $1/m^n_b$ ($n\geq 2$):
\begin{equation}
\Gamma(B\to X) = \Gamma(b\to x) \left(1+ {\cal O}(1/m_b^2)\;\right).
\end{equation}
The power-suppressed correction terms to the total inclusive widths of
both semi-- and nonleptonic decays were calculated in Refs.\ \cite{PLB,bigi}.
They depend on two hadronic matrix elements, $\lambda_1$ and
$\lambda_2$. While the latter is related to the squared mass difference
of the B and the B$^*$ meson,
\begin{equation}
\lambda_2 \approx \frac{1}{4}\,(m_{B^*}^2 - m_B^2)= 0.12\,{\rm GeV}^2,
\end{equation}
the former, $\lambda_1$, is difficult to measure; in this letter we
use $\lambda_1 =
-(0.6\pm 0.1)\,{\rm GeV}^2$ as determined from QCD sum rules \cite{BB94}.

The one-loop corrections to the partial decay widths in Eq.\ (\ref{eq:BRSL})
can be found tabulated in Ref.\ \cite{B42} except for $\Gamma(b\to ccs)$,
which was a rough estimate. In Table \ref{tab:2} we give the
corrections to this partial width. As in Ref.\ \cite{B42}, we neglect the rare
decays in the present analysis because of their smallness.
Using the same input parameters as in the last section, we find
\begin{equation}
B_{SL} = 12.0\pm 0.9 ^{+0.9}_{-1.3},\qquad \bar{B}_{SL} =
11.2\pm 1.0^{+1.0}_{-2.2}.
\end{equation}
Here the first error comes from the uncertainty in the input parameters $x_c$,
$\lambda_1$ and $\Gamma(b\to ccs)$ (in which the effect of the penguin
operators has been estimated), whereas the second one indicates the
variation of the result with the renormalization scale $\mu$. Both results
are in agreement with the most recent experimental data and the particle
data group world average $B_{SL} = (10.43\pm 0.24)\% $ \cite{data}.
Nevertheless, we observe a nonnegligible scheme-dependence of the two results.
Although at the considered order in $\alpha_s$ it is difficult to judge
which scheme is ``best'', we remark that at least for the
semileptonic width one can sum up a certain class of terms, which are of order
$\beta_0^n\alpha_s^{n+1}$. One observes that both, the explicit coefficients
multiplying $\beta_0^n\alpha_s^{n+1}$ with $n$ not too big (say $n<5$), and
the resummed all-order expression are smaller in the $\overline{\mbox{MS}}$
than in the on-shell scheme (\cite{prep}, see also \cite{summation}).
Interpreting this result with due caution, since the evidence that these
terms are dominant already in low orders comes from empirical observation
(of quantities with known complete $\alpha_s^2$ corrections) rather
than from a theoretical principle, we still feel that it favours the
$\overline{\mbox{MS}}$ scheme. Any further discussion would require
the knowledge of complete $\alpha_s^2$ or even higher
order terms, whose calculation is a formidable task.

Finally, we would like to discuss shortly the average charm quark content of
B decays, which is defined by
\begin{equation}
\langle\, n_c\,\rangle = 1 + \frac{\Gamma(B\to X_{c\bar c})}{\Gamma_{tot}}.
\end{equation}
We obtain (using again the same input parameters as in the last section):
\begin{equation}
\langle\, n_c\,\rangle = 1.27\pm 0.07,\qquad \langle\,
\bar{n}_c\,\rangle = 1.35\pm 0.19,
\end{equation}
which has to be compared with the experimental result
$\langle\, n_c\,\rangle^{exp} = 1.04\pm 0.07$ \cite{data}. The experimental
and the
theoretical numbers differ by 3 standard deviations. Unfortunately, we do
not see any natural theoretical explanation for that fact, unless $x_c$
were much smaller than we assumed, which is however in conflict with the
results obtained from the phenomenology of charmed particles.

\bigskip

{\bf Acknowledgments.} E. B.\ acknowledges the financial support of
the CYCIT, project No.\ AEN93-0520. B. F.\ and P. G.\ acknowledge gratefully
their grants from the Generalitat de Catalunya.
We thank Martin Lavelle for reading
the manuscript.

\newpage

\section*{Figure}

\begin{figure}[h]
\centerline{
\epsfysize=0.7\textheight
\epsfbox{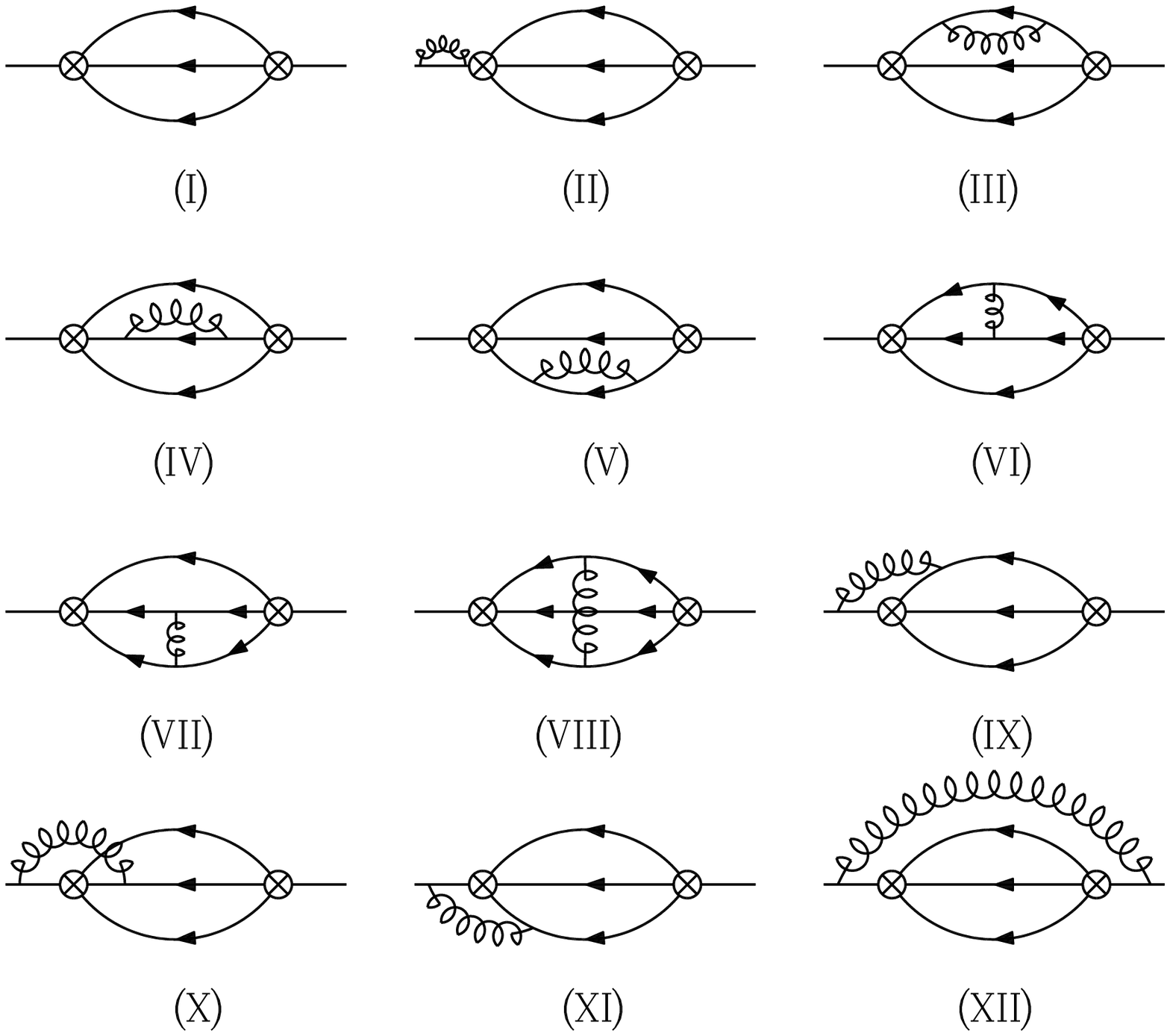}
}
\vspace{0.5in}
\caption[]{The diagrams contributing to the forward--scattering
amplitude Eq.\ (\protect{\ref{eq:FSA}}) up to order $\alpha_s$ without
penguins. The crossed circles denote insertions of any of the operators
${\cal O}_i$. Of the three internal quark lines, the upper one denotes
the c quark, the lower one the s quark, and the middle one the c
antiquark.}
\end{figure}

\clearpage

\section*{Tables}

\begin{table}[h]
\addtolength{\arraycolsep}{3pt}
\renewcommand{\arraystretch}{1.3}
$$
\begin{array}{l|llllll}
f_{ij} & 1 & 2 & 3 & 4 & 5 & 6\\ \hline
1 & 1 \\
2 & \frac{2}{3} & 1\\
3 & 2 & \frac{2}{3} & 1\\
4 & \frac{2}{3} & 2 & \frac{2}{3} & 1\\
5 & 2f & \frac{2}{3}\,f & 2f & \frac{2}{3}\,f & 1\\
6 & \frac{2}{3}\,f & 2f & \frac{2}{3}\,f & 2f & \frac{2}{3}\phantom{\,f} & 1
\end{array}
$$
\addtolength{\arraycolsep}{-3pt}
\renewcommand{\arraystretch}{1}
\caption[]{Coefficients $f_{ij}$ defined in Eq.\ (\protect{\ref{eq:deffij}}).}
\end{table}
\begin{table}[h]
\addtolength{\arraycolsep}{3pt}
\renewcommand{\arraystretch}{1.3}
$$
\begin{array}{l|cccccc}
x_c & \kappa(x_c,0\phantom{{}_s},m_b) & k_{11} & k_{12}(\mu=m_b) & k_{22} &
K(x_c,0\phantom{{}_s},m_b) & \kappa\,K\,{\rm PH}\\ \hline
0 & 1.054 & -1.34 & -7.75 & -1.41 & 1.01\pm 0.05 & 1.065\pm 0.059\\
0.1 & 1.052 & -0.14 & -6.31 & -0.53 & 1.07\pm 0.06 & 0.959\pm 0.052\\
0.2 & 1.047 & \phantom{-}2.40 & -3.50 & \phantom{-}0.99
& 1.17\pm 0.06 & 0.634\pm 0.035\\
0.3 & 1.040 & \phantom{-}6.44 & \phantom{-}0.82 & \phantom{-}2.99
& 1.29\pm 0.07 & 0.263\pm 0.015\\
0.4 & 1.032 & 14.76 & \phantom{-}9.50 & \phantom{-}5.83
& 1.45\pm 0.08 & 0.039\pm 0.002
\end{array}
$$
$$
\begin{array}{l|cccccc}
x_c & \kappa(x_c,x_s,m_b) &k_{11} & k_{12}(\mu=m_b) & k_{22} &
K(x_c,x_s,m_b) & \kappa\,K\,{\rm PH}\\ \hline
0 & 1.054 & -1.33 & -7.63 & -1.26 & 1.02\pm 0.05 & 1.062\pm 0.058\\
0.1 & 1.052 & -0.05 & -6.20 & -0.35 & 1.08\pm 0.06 & 0.956\pm 0.052\\
0.2 & 1.047 & \phantom{-}2.53 & -3.36 & \phantom{-}1.23
& 1.18\pm 0.06 & 0.631\pm 0.034\\
0.3 & 1.040 & \phantom{-}6.69 & \phantom{-}1.08 & \phantom{-}3.41
& 1.32\pm 0.07 & 0.259\pm 0.014\\
0.4 & 1.032 & 15.68 & 10.43 & \phantom{-}7.09
& 1.54\pm 0.08 & 0.037\pm 0.002
\end{array}
$$
\addtolength{\arraycolsep}{-3pt}
\renewcommand{\arraystretch}{1}
\caption[]{The LO and NLO corrections to the nonleptonic decay $b\to ccs$
as a function of $x_c=m_c/m_b$.
The penultimate column gives the increase of the decay rate
$\Gamma(b\to ccs)$ in NLO if we include finite c and s quark
effects in the radiative corrections. In the upper table we have put $m_s=0$.
The errors in $K$ represent a
conservative estimate of the unknown parts of the NLO penguin
contributions. The input parameters are $\mu=m_b=4.8\,$GeV, $x_s=0.04$ and
$\Lambda_{\overline{\rm\scriptsize MS}}^{(4)}=312\,$MeV, which corresponds to
$\alpha_s(m_Z) = 0.117$. A comparison of the last column in both tables
 shows that the effect of the strange quark mass is negligible and
actually below 5\% for all values of $x_c$.}\label{tab:2}
\end{table}
\end{document}